\documentclass[%
 preprint,
 amsmath,amssymb,
 aps,
]{revtex4-2}

\usepackage{graphicx}
\usepackage{dcolumn}
\usepackage{bm}
\usepackage{soul}
\usepackage{physics}

\begin{document}

\preprint{APS/123-QED}

\title{Spectroscopy of high pressure rubidium-noble gas mixtures}

\author{Till Ockenfels}
 \email{ockenfels@iap.uni-bonn.de}
\author{Pa\v{s}ko Roje}%
\author{Thilo vom H{\"o}vel}%
\author{Frank Vewinger}
\author{Martin Weitz}
\affiliation{%
Institut für Angewandte Physik der Universität Bonn, Wegelerstr. 8, D-53115 Bonn
}%

\date{\today}

\begin{abstract}
Spectroscopy of alkali-buffer gas mixtures at high pressures from single-digit to several 100 bars in the regime of substantial collisional broadening is relevant in a wide range of fields, ranging from collisional redistribution cooling to laboratory astrophysics. Here we report on spectroscopic measurements of dense rubidium-noble gas mixtures recorded in a pressure cell equipped with soldered sapphire optical viewports, which allows for the controlled realization of extreme conditions of high temperature and high pressure in a table top laboratory experiment. In the gas cell, we have recorded absorption and emission spectra of rubidium subject to $188\,$bar helium buffer gas pressure at $548\,$K temperature. The spectra to good accuracy follow a Boltzmann-type Kennard-Stepanov frequency scaling of the ratio of absorption and emission spectral profiles. Further, the long optical path length in the cell allowed to both record spectra of rubidium-argon mixtures at moderate temperatures and high pressures and to observe redistributional laser cooling in this system.
\end{abstract}

\maketitle

\section{Introduction}
One of the most well-proven tools for the scientific investigation of both the composition of matter and the mechanisms determining its behavior on a microscopic level is the optical spectroscopy. Over time this technique enabled the analysis of matter benefiting from the immense precision of this method. Additionally, it allows also to extract parameters of complex systems, such as temperature. In astronomy as well as in plasma physics optical spectroscopy is used as it is one of the tools to investigate the extreme conditions that the materials are exposed to, while they are out of reach for other methods of probing~\cite{pradhan2011atomic, khalafinejad2017exoplanetary,harilal2018optical}. 

In gaseous systems, when e.g.\ an atom approaches another atom, the energy levels are shifted due to the interaction, leading to a transient variation of the atom's transition frequency. On the one hand, for frequencies close to resonance with a detuning below the inverse collisional duration, the impact limit is fulfilled, and provided that the collisions are elastic a symmetric line broadening occurs, accompanied by a linear shift of the line center. On the other hand, in the spectral wings, which become especially pronounced at higher pressures due to then increased overall broadening of the line, the impact limit does not hold anymore and the lines often become asymmetric, and show a nonlinear shift with increasing pressure. A precise knowledge of the molecular potential curves is then important for a prediction of the lineshape, e.g. in the quasi-static approximation. When using inert gases as a buffer gas, the quenching of optical transitions is weak, and atoms in their electronically excited state undergo a large number of collisions, see e.g. Refs~\cite{vogl2008spectroscopy, moroshkin2014kennard, christopoulos2018rubidium, takeo1957broadening} for earlier work on the optical spectroscopy of dense alkali-noble gas mixtures. Especially the effect of a high pressure environment of the lightest noble gas helium acting as a buffer gas on the spectra of other elements is of great interest in astronomy as it is the second-most abundant element and represents 24$\%$ of the total baryonic mass in the universe~\cite{cameron1973abundances}.

In this paper we investigate absorption and emission spectra of rubidium-helium mixtures at a pressure of up to $188\,$bar and a temperature of  $548\,$K in a table top laboratory experiment.  The experiments benefit from using our proprietary high pressure cell equipped with soldered sapphire optical viewports, where the sapphire is bonded to a metal flange via active soldering making use of a compound intermediate structure allowing to mitigate thermally induced stress~\cite{ockenfels2021sapphire}. In the high density environment the frequent collisions lead to a thermalisation of the occupation distribution in the ground state and the excited state manifold. This has a number of well-known consequences for the spectral profiles: the Stokes shift and Kasha’s rule~\cite{lakowicz2013principles,schafer19731,stokes1852xxx,kasha1950characterization}, which to a certain extent are clearly present in our measured spectra. Additionally, the assumption of a thermalisation in the occupation of the excited state manifold can be tested by verifying the Kennard-Stepanov relation~\cite{kennard1918thermodynamics,stepanov1957universal} which is a Boltzmann-type scaling of the ratio of absorption and emission spectral profiles with frequency, and is known to be fulfilled in a wide range of systems like semiconductor quantum wells, doped glasses, photoactive biomolecules, dyes and other mixtures of alkali-metals and noble gases~\cite{ihara2009thermal,sawicki1996universal,croce1996excited,dau1996exciton,dobek2011influence,kawski2000local,moroshkin2014kennard,christopoulos2018rubidium}.

We report the measurement of rubidium absorption and emission spectra in the presence of high pressure helium-buffer gas. The emission spectra were recorded making use of non-resonant laser excitation, while the absorption spectra were derived from a combination of standard absorption measurements with a scaled excitation spectrum as it is described in more detail in the following. In these spectra not only the pressure broadening and shift of the resonances known as the rubidium D1- and D2-lines are clearly visible but also strong signatures of satellite resonances whose origin can be found in the form of the quasi-molecular potential curves of the collisional partners. Further it can be shown that the ratio between absorption and emission spectral profiles well follows a Kennard-Stepanov frequency scaling over a wide frequency range, however the spectral temperature extracted from the logarithmic ratio is above the cell temperature. Additionally, results on redistributional laser cooling carried out at comparable parameters but in a rubidium-argon mixture are reported.

\section{Model and theory}
\begin{figure}
\includegraphics{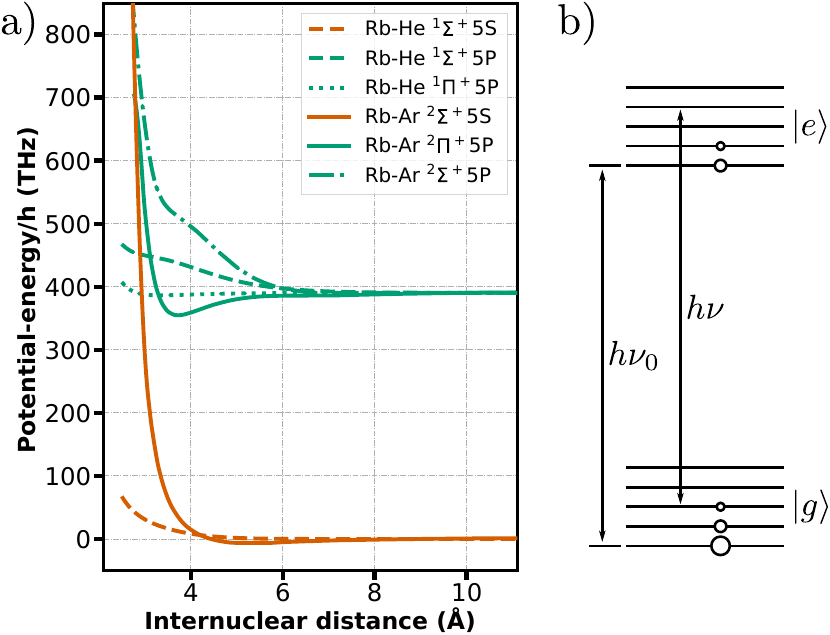}
\caption{\label{label_fig_1}a) Potential energy curves of the relevant first levels of Rb-He and Rb-Ar quasi-molecules~\cite{pascale1983use, dhiflaoui2012electronic}. b) Schematic Jablonski diagram of a two-level system with a sublevel structure in both the electronic ground and the excited states. The size of the dots is chosen so as to indicate thermal occupation in both upper and lower electronic state.}
\end{figure}

In a binary-collisional picture, the effect of the interaction of atoms is described by potential curves as depicted in Fig.~\ref{label_fig_1}a, showing the particles’ energy levels in relation to the distance to a collision partner~\cite{kuhn1934xcii,kuhn1934xciii,kuhn1937pressure}. These potentials allow for theoretical predictions of the expected spectra via calculations, e.g. based on the Franck-Condon principle when the atomic motion is so slow with respect to the collisional time such that the quasi-static approximation applies~\cite{frank1926elementary,condon1926theory}. The frequency range of the absorbed or emitted photons of a certain electronic transition is then determined by the energy differences of the potential curves, and to calculate a spectral profile weighting factors are taken into account for the relative probability of an electronic transition taking place while the perturber is at a specific distance to the atom under consideration. The quasi-static approximation is capable of predicting the shape of spectral profiles away from the resonances as well as distinct satellite resonances resulting from the form of the potential curves for a given pressure and temperature~\cite{kielkopf1976predicted}. We point out that for more accurate lineshape predictions united line broadening theories are commonly used, e.g. the Anderson-Talman method~\cite{anderson1952method, anderson1956pressure}.

Upon increased pressure, frequent collisions can furthermore lead to thermalisation within the quasi-molecular manifolds of electronic levels. For a noble buffer gas pressure in the $100\,$bar range, collisions occur on a $10^{-11}\,$s timescale, which is far faster than the upper electronic state lifetime of D-line transitions of alkali atoms. In this limit, a Boltzmann-like Kennard-Stepanov scaling between absorption and emission spectral profiles has been observed in rubidium-argon buffer gas mixtures~\cite{moroshkin2014kennard, christopoulos2018rubidium}.

Following a model developed by Sawicki and Knox~\cite{sawicki1996universal} originally in the context of the description of dye spectra, the Kennard-Stepanov relation can straightforwardly be derived for a two-level system with a ground state $\ket{g}$  and an excited state $\ket{e}$, each with an additional sublevel structure. For a long enough lifetime of the electronically excited state to allow for the sublevel structure to acquire thermal equilibrium in the presence of coupling to the thermal environment, the occupation of the sublevels in ground and excited manifolds will follow a Boltzmann distribution, as depicted in Fig.~\ref{label_fig_1}b. By making use of the Einstein A-B relation and the energy conservation, the ratio between the absorption $\alpha(\nu)$ and the emission spectra $f(\nu)$ is found to be~\cite{sawicki1996universal}: 

\begin{equation}
    \frac{f(\nu)}{\alpha(\nu)}\propto\exp \left(\frac{-h(\nu-\nu_{0})}{k_{\text{B}}T}\right)\cdot\frac{8\pi\nu^{3}}{c^{2}}\,.
    \label{label_eq_1}
\end{equation}

In the limit of the quasi-static approximation, the same result has also been derived using a standard collisional theory model~\cite{moroshkin2014kennard}. Thermalisation in electronic submanifolds here means that detailed balance applies to both the bound and free eigenstates of the quasi-molecular manifolds. For small detunings $(\nu-\nu_{0}\ll\nu_{0})$ the free space mode density $\frac{8\pi\nu^{3}}{c^{2}}$ can be assumed to be constant and formula~(\ref{label_eq_1}) can be rewritten to 

\begin{equation}
    \ln \left(\frac{\alpha(\nu)}{f(\nu)}\right)=\frac{h}{k_{\text{B}}T}\cdot\nu+D(T)\,,
    \label{label_eq_2}
\end{equation}

which shows the known linear scaling of the logarithmic ratio of the absorption and emission profile with the optical frequency.

Another consequence that follows from the relaxation within the upper electronic state manifold  is the fact that the form of the emission spectrum is independent of the excitation frequency, known as Kasha's rule. We point out that this redistribution of fluorescence is also essential for redistributional laser cooling~\cite{vogl2009laser}, as will be discussed later. 


\section{Methods / Setup}
As mentioned above, under conditions of high pressure and high temperature the shape of observed spectral profiles tends to differ significantly from the spectra observed in undisturbed systems, which makes the observation of such spectra a suitable method to investigate the conditions in these systems in a contactless way. The experiments reported in this work have been carried out in a high pressure and high temperature resistant gas cell construction recently realised by our group, providing a very reliable spectroscopic environment at these conditions. The cell is made from high-temperature stainless steel (1.4841) whose openings are enclosed by flanges which house active-soldered sapphire-windows providing optical access~\cite{ockenfels2021sapphire}. For preparation, this cell is first baked out at around $720\,$K to remove contaminations. It is then filled with 1\,g of high-purity rubidium, together with high-purity helium gas. In this filled cell we can bring the rubidium-noble gas mixture to temperatures of up to $650\,$K and noble gas pressures of up to $300\,$bars.

\begin{figure}
\includegraphics{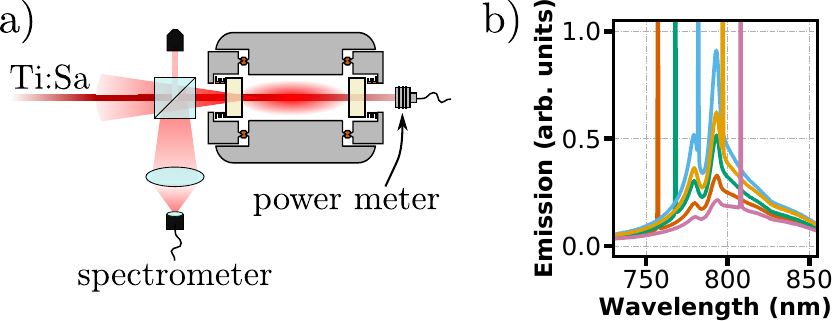}
\caption{\label{label_fig_2}a) Optical setup for the detection of the emission from the gas mixture in a cell with one optical axis. The wavelength-tunable excitation laser beam is coupled into the cell through a polarizing beam splitter (PBS) where it excites atoms in the gas. The part of the resulting emission leaving the cell through the entrance window and reflected by the PBS is focused into a fibre coupler and sent to the spectrometer. b) Emission spectra of the rubidium-helium mixture recorded for different excitation frequencies. As the shape of the spectral profile of the quasi-molecular emission is always the same, Kasha's rule is well fulfilled.}
\end{figure}

Figure~\ref{label_fig_2}a shows the setup we used to record the emission spectra of the gas mixture. We make use of a tunable titanium-sapphire laser, providing an optical power above $\,$1W in the wavelength range from $715$ to $950\,$nm. With this laser we excite the system close to the rubidium D-lines. Emitted fluorescence from the gas sample transmitted back through the entrance window and redirected by a beam splitter cube is focused into a multimode fiber attached to an optical spectrometer. We record spectra at different excitation  wavelengths (see Fig.~2b), and the observed spectral signals, apart from showing scattered excitation beam light around the excitation wavelength, clearly follow Kasha’s rule. Correspondingly, we combine multiple spectra at different excitation  wavelengths to obtain emission spectra without residual excitation light.

Absorption spectra of the dense gas mixture are recorded by determining the variation of the fluorescence yield detected in backwards direction as a function of laser wavelength, which in the case of redistribution can well be assumed to be proportional to the absorption~\cite{christopoulos2018rubidium}. This yields raw data as in Fig.~\ref{label_fig_2}b, note the visible variation of the line area for different incident laser wavelengths, the latter can be read off from the peak caused by scattered excitation light. To find the absorption coefficient we scale the data obtained from the wavelength dependence of the integrated fluorescence (naturally excluding the peak caused by scattered laser light) by overlapping them with data measured far from resonance at low optical density, where one can determine the absorption coefficient by directly comparing incident and transmitted beam power. 


To test for an induced temperature change in the gas mixture due to redistributional laser cooling we use the method of deflection spectroscopy, where a position-dependent deflection of a second, non-resonant laser beam is recorded. From this deflection profile the temperature profile in the cell can be determined via contactless probing. 

\section{Results of rubidium-helium gas mixture spectroscopy}
\begin{figure}
\includegraphics{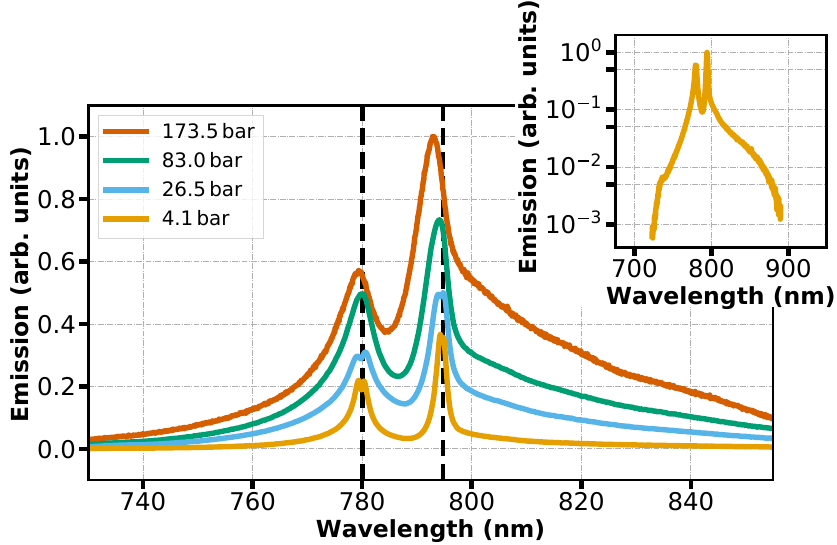}
\caption{\label{label_fig_3}Emission spectra from the rubidium-helium gas mixture measured at a cell temperature of $507\,$K and different pressures. The two vertical dashed lines mark the position of the undisturbed D1- and D2-resonances. A pressure-dependent broadening and blue-shift is visible. Inset: The emission spectrum at $4.1\,$bar shown in logarithmic scale for better comparison with the data in~\cite{bouhadjar2014rubidium} and \cite{allard2006collisional}.}
\end{figure}

In Fig.~\ref{label_fig_3} emission spectra of rubidium-helium mixtures are shown recorded at a gas cell temperature of $507\,$K and a wide range of pressures from 4.1 to $173.5\,$bar. The rubidium density is assumed to be vapor-limited, at a partial pressure of $~0.22\,$mbar at this temperature. One sees strong broadening of the D1- and D2-resonances with pressure. Also, a pressure shift of both emission maxima from the position of the undisturbed resonances is clearly visible. The mean shift can be extracted from the spectra via a Lorentzian fit to the maxima to be $(4.82\pm0.07)\,$GHz/bar for the D1- and $(2.01\pm0.12)\,$GHz/bar for the D2-line, respectively. For the D1-transition this is around 40\% larger than the literature values as they are presented in Tab.~\ref{table1}, however we want to point out that the here determined shift of the emission line can differ from the shift of the absorption line, especially in light of the thermal redistribution present at high pressures. Also earlier work on emission spectra \cite{ottinger1975broadening} gives different line shifts compared  to those for absorption profiles.

For the D2-line a much smaller pressure shift has been reported in earlier work, however in a different pressure range, only the measurement at highest pressure so far by Shang-Yi also shows a relatively strong shift of the D2-line. We attribute this significant difference to the other literature values mainly to the high pressures used in the presented work, where the assumption of a purely linear shift with pressure is not valid, as e.g. a thermalisation within the electronically excited D-lines quasi-molecular rovibrational manifolds occurs. Evidence for the thermalization present for the observed spectra is the relative line strength of the D1- and D2-lines, which is altered from a $1:2$ ratio expected from the multiplicity, e.g.\ observed in the low pressure regime, while here the lower energetic D1-line emission is favored.

\begin{table*}[]
\caption{\label{table1}
Overview of pressure-induced shifts of absorption and emission lines of the rubidium principal series reported in earlier and this publication.}
\begin{ruledtabular}

\begin{tabular}{l ccc c cccc}
Publication &    year  & $P_{\text{max}}\,$[bar] & $T\,$[K] & \multicolumn{2}{c}{ D1-shift$\,$[Ghz/bar]} & \multicolumn{2}{c}{ D2-shift$\,$[Ghz/bar]} \\\cline{5-6}\cline{7-8}
&&&& Abs.&  Em. \hfill&Abs.&Em.\\
\hline
Shang-Yi~\cite{shang1940broadening}          & 1940 & 101             & 580        &3.33 &       & 1.34&   \\
Ottinger~\cite{ottinger1975broadening}     & 1974 & 1.5             & 318          &   &        & & 0.52            \\
Romalis~\cite{romalis1997pressure}          & 1997 & 13              & 353        & 3.3   &     & 0.35    &          \\
Rotondaro~\cite{rotondaro1997collisional}          & 1997 & 0.4             & 394        & 3.62    &          & 0.28  &          \\
Miller~\cite{miller2016high}      & 2016 & 20              & 343        & 3.54     &          & 0.15   &           \\
This publication               &      & 173             & 507          & &4.82             & &2.01            
\end{tabular}
\end{ruledtabular}
\end{table*}

In the inset in Fig.~\ref{label_fig_3} the emission spectrum of the rubidium-helium mixture recorded for the lowest pressure of $4.1\,$bar is shown again in logarithmic scale. In this visualisation one sees that the form of the spectrum away from the resonance is not the same for red- and blue-detuning. Red-detuned from the resonances the emission strength drops off relatively slowly and for wavelengths higher than  $865\,$nm the slope changes and a shoulder-like feature arises. For blue-detuning the emission decreases much faster, yet around $735\,$nm an additional feature in the form of a satellite resonance is visible. In the literature predicted spectra for this system at slightly different parameters (temperature of $1000\,$K and a density of $1\times 10^{20}\,$cm$^{-3}$, equating to around $14\,$bar at that temperature) can be found~\cite{bouhadjar2014rubidium,allard2006collisional}. When comparing the general form of the emission spectra they are nevertheless in good agreement. The shape of the red wing with its changing slope is present in the calculated spectra as well as the satellite peak at blue detuning. 

\begin{figure}
\includegraphics{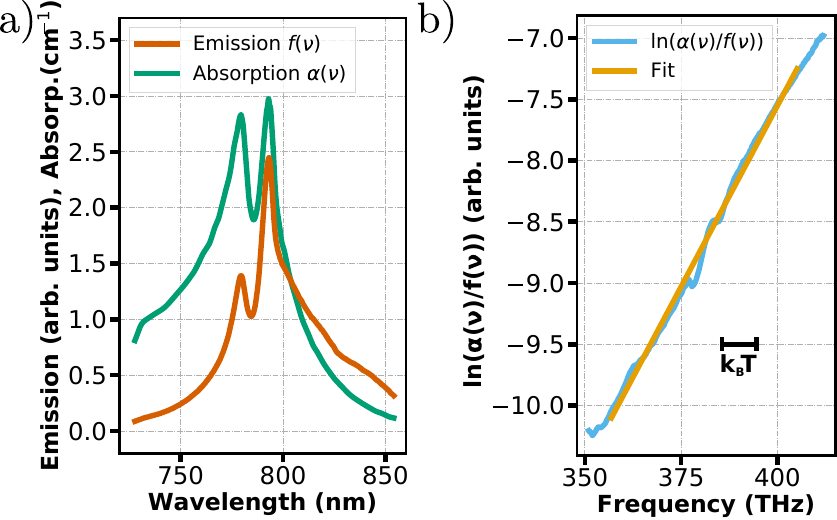}
\caption{\label{label_fig_4}a) Emission and absorption spectra of the Rb-He mixture measured at $548\,$K and $188\,$bar (In comparison to Fig.~\ref{label_fig_3} a higher temperature was chosen to be able to observe the satellite resonance). b) The natural logarithm of the ratio of the spectral profiles versus the optical frequency.} 
\end{figure}

In Fig.~\ref{label_fig_4}a measured absorption as well as emission spectra of the rubidium-helium system are shown for a pressure of $188\,$bar. The use of such high pressures well ensures a full redistribution of fluorescence, as is necessary to apply the above described method (see chapter 3) of extracting the absorption strength from the detected fluorescence yield, as this method relies heavily on the spectral profile of the emission being independent of the excitation frequency. The absorption shows a smooth decline in the absorption strength red-detuned to the resonances while in the blue wing a satellite resonance at $865\,$nm is visible, which becomes visible at the same position in the calculated spectra in~\cite{bouhadjar2014rubidium} for temperatures of $320\,$K and larger. This satellite is of certain interest for the realization of exciplex-pumped alkali-gas lasers~\cite{RICE2019550}. 

To verify the before claimed fulfillment of the Kennard-Stepanov relation in Fig.~\ref{label_fig_4}b the natural logarithm of the ratio of the absorption and emission spectra is shown. This ratio is in very good agreement with the plotted linear function over a large range of $100\,$nm, which equates to $5.9\,k_{\text{B}}T$ (compare formula~(\ref{label_eq_2})). Small discrepancies between the logarithmic ratio and the linear fit can be seen at the positions of the resonances which is attributed to the maxima not being fully resolved due to technical limitations. Also for a large red-detuning from the resonances the emission signal shows a satellite peak where there is no additional feature in the absorption spectrum which leads to a disturbance in the logarithmic ratio. The slope of the fitted function corresponds to a spectral temperature of $812\,$K, which is above the cell temperature of $548\,$K. The discrepancy is most probably 
either due to an incomplete thermalization of the upper electronic state quasimolecular manifold at the used pressure for the light helium buffer gas or a finite quantum yield of the system in the presence of this noble gas. To investigate the origin of the incomplete thermalization in more detail, time-resolved spectroscopy could be carried out in future. In earlier work of our group studying mixtures of alkalis with the heavier argon noble gas, the spectral temperature was in very good agreement with the cell temperature~\cite{christopoulos2017verifying,christopoulos2018rubidium}.

\section{Results of rubidium-argon gas mixture spectroscopy}
\begin{figure*}
\includegraphics{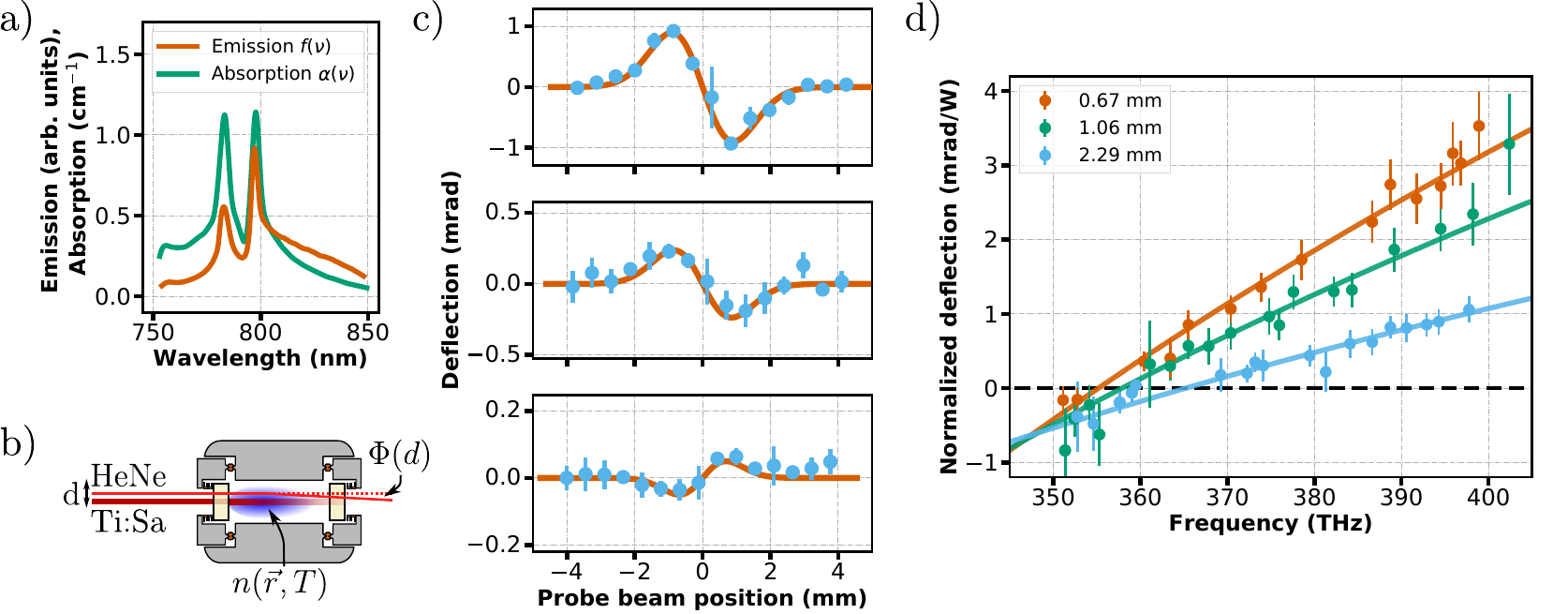}
\caption{\label{label_fig_5}a) Emission and absorption spectra of the Rb-Ar mixture measured at $530\,$K and $131\,$bar. b)
Sketch of the used optical setup for measurements of the temperature-induced deflection of the probe beam. c) Individual deflection signals at three different cooling beam frequencies ($402.3\,$THz ($745.0\,$nm), $365.4\,$THz ($820.2\,$nm), and $352.6\,$THz ($849.8\,$nm)). The orientation being flipped for the most red-detuned excitation indicates cooling of the gas. Recorded with a beam diameter of the cooling laser of $1.06$\,mm. d) Normalised deflection measurements for different cooling beam diameters together with fitted functions.}
\end{figure*}

While the optical path length of $100\,$mm of the cell described here is obstructive when trying to resolve the maximum absorption strength on resonance it yet enables measurements on gas mixtures of relatively low temperature resulting in little rubidium evaporation and therefore relatively small rubidium densities of $~7\cdot10^{10}\,$cm$^{-3}$, corresponding to a partial pressure of $0.5\,$mbar. In Fig.~\ref{label_fig_5}a emission and absorption spectra of a rubidium-argon mixture at a temperature of $530\,$K and $131\,$bar of total pressure are shown. Due to the full redistribution of the fluorescence the form of the emission spectrum here to good accuracy is not dependent on the excitation frequency. In comparison with earlier publications~\cite{moroshkin2014kennard,christopoulos2018rubidium} these measurements were carried out at significantly different cell temperatures and pressures respectively, which results in altered spectral profiles.

Further, the greater optical path length also allows for a sufficient interaction between the photons from the excitation laser beam and the gas mixture such that even for these relatively low rubidium densities a relative cooling of a macroscopic amount of gas can be observed when a sufficiently red-detuned excitation frequency for the excitation is chosen. This so called redistributional laser cooling was first reported in~\cite{vogl2009laser}, and is a result of the mean fluorescence of the rubidium atoms being shifted from the excitation frequencies, as the emission profile follows Kasha’s rule due to the frequent collisons of the excited rubidium atoms with buffer gas atoms during their electronic state lifetime. By excitation with a red-detuned laser beam this allows for the extraction of energy from the system, as described by~\cite{vogl2009laser}:

\begin{eqnarray}
    P_{\text{cool}}(\nu_{\text{L}})&=&P_{\text{in}}(\nu_{\text{L}})\cdot\alpha(\nu_{\text{L}})\cdot\frac{\nu_{\text{fl}}-\nu_{\text{L}}}{\nu_{\text{fl}}}\nonumber\\
    &=&P_{\text{abs}}(\nu_{\text{L}})\cdot\frac{\nu_{\text{fl}}-\nu_{\text{L}}}{\nu_{\text{fl}}}\,.
    \label{label_eq_3}
\end{eqnarray}

This expected cooling power is determined by the incoming beam power $P_{\text{in}}(\nu_{\text{L}})$, the absorption coefficient for the particular excitation frequency $\alpha(\nu_{\text{L}})$ (determining the interaction strength), and the detuning factor, which is a measure for the relative energy difference between the absorbed and the emitted photons. The detuning factor determines how much energy is extracted in one cooling cycle.

The resulting temperature change in the gas can be measured via the angular deflection of a second, off-resonant probe laser beam (HeNe-laser with a wavelength of 633$\,$nm$\,($473$\,$THz), a power of $2\,$mW and a beam diameter of $0.5\,$mm) for different relative distances to the excitation laser beam (for a sketch of the setup see Fig.~\ref{label_fig_5}b), as it is shown in Fig.~\ref{label_fig_5}c for three excitation frequencies. As described in~\cite{vogl2009laser} in more detail, the deflection results from the temperature change in the gas which in turn according to the corresponding local change of the gas density leads to a change in the refractive index and thereby to the deflection of the probe beam. The polarity of this signal indicates whether the gas was heated up or cooled down during the cycles of excitation by the laser photons and the subsequent emission of frequency-shifted photons. As can be seen for the lowest-energetic excitation the deflection signal is flipped as the gas is cooled. Fig.~\ref{label_fig_5}d shows corresponding data for the deflection normalized to the absorbed power ($P_{\text{abs}}(\nu_{\text{L}})= P_{\text{in}}(\nu_{\text{L}})\cdot\alpha(\nu_{\text{L}})$) for three different cooling beam diameters, which within experimental accuracy follows the expected linear scaling. The observed magnitude of the slope reduces with increasing beam diameter, as can be understood from the smaller gradient of the imprinted temperature profile.


In a perfect system the change of sign in the deflection (which indicated the transition from heating to cooling) should take place when the excitation frequency becomes smaller than the mean fluorescence frequency, as then energy is extracted from the system. But, as pointed out already in~\cite{vogl2009laser}, this only holds if no other effects, e.g.\ finite quantum efficiency are present which can lead to energy being deposited in the system, counteracting the redistributional cooling. This would result in a shift of this crossover point towards smaller excitation frequencies, as indeed is visible in   Fig.~\ref{label_fig_5}d. As the observed crossover point shifts to lower frequencies for smaller beam diameters, it seems that loss processes depend nonlinearly on intensity, as could well be explained by excitation to higher lying electronic rubidium states by energy pooling in the dense gas mixture~\cite{christopoulos2017verifying,allegrini1977molecule}.

\section{Conclusions}
To conclude, we have reported absorption and emission spectroscopic measurements of rubidium atoms subject to helium buffer gas pressures up to $188\,$bar range at elevated temperatures. Spectra recorded in the presence of this light noble gas, which is of astrophysical relevance, show a Boltzmann-type Kennard-Stepanov frequency scaling between absorption and emission, thus extending measurements previously carried out in dense gas mixtures with heavier noble gases. Other than in the case of alkali spectra recorded in the presence of the heavier argon noble gas, the here observed spectral temperature is above the cell temperature. We have also reported spectroscopy and redistribution cooling measurements on rubidium-argon gas mixtures at comparable temperature and pressure conditions, which extend earlier works on redistribution laser cooling. The here observed difference in spectral and thermodynamic temperature for the light helium buffergas rubidium spectra hints at either incomplete thermalization of the quasimolecular manifold or a lower quantum efficiency than in the rubidium argon system at comparable buffer gas pressures. To characterize thermalization time scales for the present system as well as to identify potential loss or quenching channels, it would be interesting to do time-resolved studies of the spectra for different noble gas rubidium mixtures.
For the future, it will also be interesting to test for evidence of possible thermalization effects of quasi-molecular manifolds in the spectra of dense astrophysical objects. Laboratory experiments at elevated pressure and temperature conditions, as required to allow for critical confrontations of spectroscopic theory with experimental studies of dense gases, demand for technically challenging cell construction. 
Additionally, for future studies of redistribution laser cooling, it would be helpful to find spectroscopic samples that provide sufficient optical density at room temperature conditions. Possible candidate systems are e.g. mixtures containing acetylene or formaldehyde molecules, which have electronic transitions in the ultraviolet $200-300\,$nm wavelength regime accessible to frequency converted cw-laser sources, and a noble buffer gas.

\begin{acknowledgments}
We acknowledge support from the DFG (Grant Nos. WE 1748-15, 581412 and SFB/TR 185, 277625399)
\end{acknowledgments}

\bibliography{apssamp}

\end{document}